\documentclass[onecolumn]{aa}
\usepackage{latexsym,graphicx,amssymb,verbatim,float,natbib,amsmath}
\usepackage{hyperref}
\usepackage{txfonts}
\begin{document}
\title{GeV-TeV cosmic-ray spectral anomaly as due to re-acceleration by weak shocks in the Galaxy}
\author{Satyendra Thoudam\thanks{E-mail: s.thoudam@astro.ru.nl} and J\"{o}rg R. H\"{o}randel}
\institute{Department of Astrophysics, IMAPP, Radboud University Nijmegen\\P.O. Box 9010, 6500 GL Nijmegen, The Netherlands\\
}
\date{\today}

\abstract{
Recent cosmic-ray measurements have found an anomaly in the cosmic-ray energy spectrum at GeV-TeV energies. Although the origin of the anomaly is not clearly understood, suggested explanations include effect of cosmic-ray source spectrum, propagation effects, and the effect of nearby sources. In this paper, we propose that the spectral anomaly might be an effect of re-acceleration of cosmic rays by weak shocks in the Galaxy. After acceleration by strong supernova remnant shock waves, cosmic rays undergo diffusive propagation through the Galaxy. During the propagation, cosmic rays may again encounter expanding supernova remnant shock waves, and get re-accelerated. As the probability of encountering old supernova remnants is expected to be larger than the younger ones due to their bigger sizes, re-acceleration is expected to be mainly due to weaker shocks. Since weaker shocks generate a softer particle spectrum, the resulting re-accelerated component will have a spectrum steeper than the initial cosmic-ray source spectrum produced by strong shocks. For a reasonable set of model parameters, it is shown that such re-accelerated component can dominate the GeV energy region while the non-reaccelerated component dominates at higher energies, explaining the observed GeV-TeV spectral anomaly.
}
\authorrunning{Thoudam and H\"{o}randel}
\titlerunning{GeV-TeV cosmic-ray spectral anomaly as due to re-acceleration by weak shocks in the Galaxy}
\maketitle

\section{Introduction} 
Measurements of cosmic rays by the ATIC \citep{Panov2007}, CREAM \citep{Yoon2011}, and PAMELA \citep{Adriani2011} experiments have found a spectral anomaly at GeV-TeV energies. The spectrum in the TeV region is found to be harder than at  GeV energies. Although the hardening is found to be more prominent in the proton and helium spectra, it also seems to be present in the spectra of heavier cosmic-ray elements, such as carbon and oxygen. The spectral anomaly is difficult to explain using simple general models of cosmic-ray acceleration, and their transport in the Galaxy. Simple linear theory of cosmic-ray acceleration \citep{Krymskii1977, Bell1978, Blandford1978}, and the nature of their propagation in the Galaxy \citep{Ginzburg1976} predict a single power-law cosmic-ray spectrum over a wide range in energy.

The origin of the anomaly is still not clearly understood. Possible explanations that have been suggested include the effect of cosmic-ray source spectrum \citep{Biermann2010, Ohira2011, Yuan2011, Ptuskin2013}, effects due to propagation through the Galaxy, \citep{Tomassetti2012, Blasi2012, Aloisio2013}, and the effect of nearby sources (\citealp{Thoudam2012}, \citealp{Thoudam2013}; \citealp{Erlykin2012}; \citealp{Bernard2013}; \citealp{Zatsepin2013}). 

In this paper, we discuss the possibility that the anomaly can be an effect of re-acceleration of cosmic rays by weak shocks in the Galaxy. This scenario was also shortly discussed recently by \citealp{Ptuskin2011}. After acceleration by strong supernova remnant shock waves, cosmic rays escape from the remnants and undergo diffusive propagation in the Galaxy. The propagation can be accompanied by some level of re-acceleration due to repeated encounters with expanding supernova remnant shock waves \citep{Wandel1988, Berezhko2003}. As older remnants occupy a larger volume in the Galaxy, cosmic rays are expected to encounter older remnants more often than the younger ones. Thus, this process of  re-acceleration is expected to be produced mainly by weaker shocks. As weaker shocks generate a softer particle spectrum, the resulting re-accelerated component will have a spectrum steeper than the initial cosmic-ray source spectrum produced by strong shocks. As will be shown later, the re-accelerated component can dominate at GeV energies, while the non-reaccelerated component (hereafter referred to as the ``normal component") dominates at higher energies.

Cosmic rays can also be re-accelerated by the same magnetic turbulence responsible for their scattering and spatial diffusion in the Galaxy. This process, which is commonly known as the distributed re-acceleration, has been studied quite extensively, and it is known that it can produce strong features on some of the observed properties of cosmic rays at low energies. For instance, the peak in the secondary-to-primary ratios at $\sim 1$ GeV/nucleon can be attributed to this effect \citep{Seo1994}. Earlier studies suggest that a strong amount of re-acceleration of this kind can produce unwanted bumps in the cosmic-ray proton and helium spectra at few GeV/nucleon \citep{Cesarsky1987, Stephens1990}. However, it was later shown that for some mild re-acceleration which is sufficient to reproduce the observed boron-to-carbon ratio, the resulting proton spectrum does not show any noticeable bumpy structures \citep{Seo1994}. In fact, the efficiency of distributed re-acceleration is expected to decrease with energy, and its effect becomes negligible at energies above $\sim 20$ GeV/nucleon.

On the other hand, for the case of encounters with old supernova remnants mentioned earlier, the re-acceleration efficiency does not depend significantly on the energy. It depends mainly only on the rate of supernova explosions, and the fractional volume occupied by supernova remnants in the Galaxy. Hence, its effect can be extended to higher energies compared to that of the distributed re-acceleration, as also noted in \citealp{Ptuskin2011}. As in the case of distributed re-acceleration, this kind of re-acceleration will also be strongly constraint by the measured secondary-to-primary ratios. In the present study, we will first determine the maximum amount of re-acceleration permitted by the available measurements on the boron-to-carbon ratio. Then, applying  the same strength of re-acceleration to the proton and helium nuclei, we will show for a reasonable set of model parameters that this kind of re-acceleration can be responsible for the observed spectral hardening.  

\section{Transport equation with re-acceleration}
Following \citealp{Wandel1987}, the re-acceleration of cosmic rays in the Galaxy is incorporated in the cosmic-ray transport equation as an additional source term with a power-law spectrum. Then, the steady-state transport equation for cosmic-ray nuclei undergoing diffusion, re-acceleration and interaction losses can be written as,
\begin{equation}
\nabla\cdot(D\nabla N)-\left[\bar{n} v\sigma+\xi\right]\delta(z)N+\left[\xi sp^{-s}\int^p_{p_0}du\;N(u)u^{s-1}\right]\delta(z)=-Q\delta(z)
\end{equation}
where we use cylindrical spatial coordinates with the radial and vertical distance represented by $r$ and $z$ respectively, $p$ is the momentum/nucleon of the nuclei, $N(\textbf{r},p)$ represents the differential number density, $D(p)$ is the diffusion coefficient, and $Q(r,p)\delta(z)$ represents the rate of injection of cosmic rays per unit volume by the source. The first term in Eq. (1) represents diffusion. The second term represents losses due to inelastic interactions with the interstellar matter, and also due to re-acceleration to higher energies, where $\bar{n}$ represents the averaged surface density of interstellar atoms, $v(p)$ the particle velocity, $\sigma(p)$ the inelastic collision cross-section, and $\xi$ corresponds to the rate of re-acceleration. The third term with the integral represents the generation of particles via re-acceleration of lower energy particles. It assumes that a given cosmic-ray population is instantaneously re-accelerated to form a power-law distribution with an index $s$. Eq. (1) does not include ionization losses and the effect of convection due to the Galactic wind which are important mostly at energies below $1$ GeV/nucleon. In pure diffusion model, these processes can be safely neglected above $1$ GeV/nucleon. But, in re-acceleration model, these processes (in particular the ionization losses) can affect the result at high energies because the number density of re-accelerated cosmic rays depends on the density of low-energy particles. Including ionization losses will reduce the number of low-energy particles available for re-acceleration than in the case without ionization. Comparing analytical solution without ionization losses with the result obtained from numerical calculation that incorporate ionization, \citealp{Wandel1987} had shown that the ionization effect can be reproduced quite well by truncating the particle distribution at a certain low energy at approximately $100$ MeV/nucleon. In our calculation, we introduce such a low-energy cut-off to approximate the effect due to losses at low energies.

The cosmic-ray propagation region is assumed to be a cylindrical region, bounded in the vertical direction at $z=\pm H$, and unbounded in the radial direction. Both the matter and the source are assumed to be uniformly distributed in an infinitely thin disk of radius $R$ located on the Galactic disk $(z=0)$. This assumption is based on the known high concentration of supernova remnants, and atomic and molecular hydrogens near the Galactic disk. For cosmic-ray primaries, the source term $Q(r,p)$ is thus taken as $Q(r,p)=\bar{\nu} H[R-r]H[p-p_0]Q(p)$, where $\bar{\nu}$ denotes the rate of supernova explosions (SNe) per unit surface area on the disk, $H(m)=1 (0)$ for $m>0 (<0)$ represents the Heaviside step function, and $p_0$ (which also serves as the lower limit in the integral in Eq. 1) is the low-momentum cut-off we have introduced to approximate the ionization losses and is chosen to correspond to a kinetic energy of $100$ MeV/nucleon. The source spectrum $Q(p)$ is assumed to follow a power-law in total momentum with a high-momentum exponential cut-off. In terms of momentum/nucleon, it can be expressed as 
\begin{equation}
Q(p)=AQ_0 (Ap)^{-q}\exp\left(-\frac{Ap}{Zp_c}\right)
\end{equation}
where $A$ and $Z$ represents the mass number and charge of the nuclei respectively, $Q_0$ is a constant related to the amount of energy $f$ injected into a cosmic ray species by a single supernova event, $q$ is the source spectral index which is taken to be always less than the re-accelerated index $s$, and $p_c$ is the high-momentum cut-off for protons. In writing Eq. (2), we assume that the maximum total momentum (or energy) for a cosmic-ray nuclei produced by a supernova remnant is $Z$ times that of the protons. Moreover, the diffusion coefficient as a function of particle rigidity is assumed to follow  $D(\rho)=D_0\beta(\rho/\rho_0)^a$, where $\rho=Apc/Ze$ is the particle rigidity, $D_0$ is the diffusion constant, $\beta=v/c$ with $c$ the velocity of light, $a$ is the diffusion index, and $\rho_0$ is a constant.

In the present model, since the re-acceleration of cosmic rays is considered to be produced by their encounters with  supernova remnants, it follows that re-acceleration occurs only in the Galactic disk. The rate of re-acceleration depends on the rate of supernova explosions and the fractional volume occupied by supernova remnants in the Galaxy. If $V=4\pi \Re^3/3$ is the volume occupied by a supernova remnant of radius $\Re$, then in Eq. (1), $\xi=\eta V\bar{\nu}$, where $\eta$ is a correction factor for $V$ we have introduced to take care of the unknown actual volume of the supernova remnants that re-accelerate cosmic rays. For the present study, we keep $\eta$ as a parameter which will be  determined later based on the observed cosmic-ray data, and we take $\Re=100$ pc which is roughly the typical radius of a supernova remnant of age $10^5$ yr expanding in the interstellar medium with an initial shock velocity of $10^9$ cm s$^{-1}$. 
  
The solution of Eq. (1) can be obtained by solving the transport equation separately for the regions above and below the Galactic disk ($z>0$ and $z<0$ respectively), and by connecting the two solutions through the flux continuity relation at $z=0$. We use the standard Green's function technique in solving Eq. (1). The Green's function is obtained to be (see Appendix A for the derivation)
\begin{equation}
G(\textbf{r},\textbf{r}^\prime,p,p^\prime)=\frac{1}{2\pi}\int^\infty_0 kdk\; F(p,p^\prime)\times\frac{\sinh\left[k(H-z)\right]}{\sinh(kH)}\times \mathrm{J_0}\left[k(r-r^\prime)\right]
\end{equation}
where $\mathrm{J_0}$ is a Bessel function of order 0, and
\begin{align}
&F(p,p^\prime)=\frac{1}{L(p)}\left[\delta(p-p^\prime)+H[p-p^\prime]\frac{\xi s {p^\prime}^{s-1}}{p^s L({p^\prime})}\times \exp\left(\xi s\int^{p^\prime}_p I(u) du\right)\right]\nonumber\\
& I(u)=-\frac{1}{uL(u)}
\end{align}
with the function $L$ defined by,
\begin{equation}
L(m)=2D(m)k\coth(kH)+\bar{n}v(m)\sigma(m)+\xi
\end{equation}

In deriving Eq. (3), we have already incorporated the assumption that the sources are distributed in the Galactic disk, thus no $z^\prime$ is appearing in the equation. Assuming azimuthal symmetry for the source distribution, the cosmic-ray density at a position $\textbf{r}$ is obtained as
\begin{align}
N(\textbf{r},p)&=2\pi\int^{\infty}_{0}dp^\prime\int^{\infty}_0r^\prime dr^\prime G(\textbf{r},\textbf{r}^\prime,p,p^\prime) Q(r^\prime,p^\prime)\nonumber\\
&=2\pi\bar{\nu}\int^{\infty}_{0}dp^\prime\int^{\infty}_0r^\prime dr^\prime G(\textbf{r},\textbf{r}^\prime,p,p^\prime) H[R-r^\prime]H[p^\prime-p_0]Q(p^\prime)\nonumber\\
&=2\pi\bar{\nu}\int^{\infty}_{p_0}dp^\prime\int^R_0r^\prime dr^\prime G(\textbf{r},\textbf{r}^\prime,p,p^\prime)Q(p^\prime)
\end{align}

Substituting for $Q(p^\prime)$ from Eq. (2) and the Green's function in Eq. (6), and performing the integral over $r^\prime$ and also the $p^\prime$ integral involving the delta function, the density of cosmic-ray primaries at $r=0$ is obtained as,
\begin{align}
N(z,p)&=\bar{\nu} R\int^{\infty}_0 dk\; \frac{\sinh\left[k(H-z)\right]}{\sinh(kH)}\times \frac{\mathrm{J_1}(kR)}{L(p)}
\times\left\lbrace H[p-p_0]Q(p)+\xi sp^{-s}\int^{\infty}_{p_0}H[p-p^\prime]{p^\prime}^s dp^\prime\;Q(p^\prime)\mathcal{A}(p^\prime)\times\exp\left(\xi s\int^p_{p^\prime} \mathcal{A}(u)du\right)\right\rbrace\nonumber 
\end{align}
where $\mathrm{J_1}$ is a Bessel function of order 1, and the function $\mathcal{A}=-I$ where $I$ is given by Eq. (4). For cosmic rays with $p>p_0$, $H[p-p_0]$ in the above equation can be set to $1$. Moreover, as $H[p-p^\prime]$ is non-zero only for $p>p^\prime$, the upper limit in the integral $\int^{\infty}_{p_0}dp^\prime$ can be replaced by $p$ and set $H[p-p^\prime]$ also to $1$. Then, the primary cosmic-ray density at $r=0$ for $p>p_0$ can be written as,
\begin{align}
N(z,p)&=\bar{\nu} R\int^{\infty}_0 dk\; \frac{\sinh\left[k(H-z)\right]}{\sinh(kH)}\times \frac{\mathrm{J_1}(kR)}{L(p)}
\times\left\lbrace Q(p)+\xi sp^{-s}\int^p_{p_0}{p^\prime}^s dp^\prime\;Q(p^\prime)\mathcal{A}(p^\prime)\times\exp\left(\xi s\int^p_{p^\prime} \mathcal{A}(u)du\right)\right\rbrace 
\end{align}

Considering that the position of our Sun is very close to the Galactic plane, the cosmic-ray density at the Earth can be calculated from Eq. (7) taking $z=0$. The first term within the curly bracket on the right hand side of Eq. (7) is the normal cosmic-ray component, which has not suffered re-acceleration, and the second term is purely the re-accelerated component. For a given diffusion index, the high-energy spectra of the two components are shaped by their respective injection indices $q$ and $s$, and their spectral indices approach $q+a$ and $s+a$ respectively. As re-acceleration takes away particles from the low-energy region and feeds them into the higher energy part of the spectrum, for re-acceleration by weak shocks for which $s>q$, the re-accelerated component might become visible as a bump or enhancement in the energy spectrum at a certain energy range. In the case of re-acceleration by strong shocks which produces a harder particle spectrum, say $s=q$, the effect of re-acceleration will be hard to notice as both the components will have the same spectra in the Galaxy. These have been extensively discussed in \citealp{Wandel1987}.  

For cosmic-ray secondaries, their equilibrium density in the Galaxy can be obtained following a similar procedure to their primaries described above, but with the source replaced by
\begin{equation}
Q_2(\textbf{r},p)=\bar{n} v_1(p)\sigma_{12}(p)H[R-r]H[p-p_0]N_1(\textbf{r},p) \delta(z)
\end{equation}
where $v_1$ represents the velocity of the primary nuclei, $\sigma_{12}$ represents the total fragmentation cross-section of the primary to the secondary, and $N_1$ is the primary nuclei density given by Eq. (6). The subscripts $1$ and $2$ have been introduced to denote the primary and secondary nuclei respectively. The secondary cosmic-ray density at $r=0$ for $p>p_0$ is given by
\begin{align}
N_2(z,p)&=R\int^{\infty}_0 dk\; \frac{\sinh\left[k(H-z)\right]}{\sinh(kH)}\times \frac{\mathrm{J_1}(kR)}{L_2(p)}\times\left\lbrace Q_2(0,p)+\xi sp^{-s}\int^p_{p_0}{p^\prime}^s dp^\prime\;Q_2(0,p^\prime)\mathcal{A}_2(p^\prime)\times\exp\left(\xi s\int^p_{p^\prime} \mathcal{A}_2(u)du\right)\right\rbrace 
\end{align}
where $L_2$ follows the same definition as given by Eq. (5), but with all quantities referring to the secondary nuclei. The source term in Eq. (9) $Q_2(0,m)=\bar{n} v_1(m)\sigma_{12}(m)N_1(0,m)$, where $N_1(0,m)$ represents the primary density at $z=0$ which can be calculated using Eq. (7), and $\mathcal{A}_2=-I_2$ with $I_2$ defined as given by Eq. (4) with $L$ replaced by $L_2$. In Eq. (9), the first term on the right hand side represents secondary cosmic rays which have not been re-accelerated after their production in the interstellar medium, and the second term represents those which have undergone re-acceleration. It is important to realize that the second term contains a re-accelerated component which is produced by the re-accelerated primaries. This component can lead to stronger re-acceleration signatures on the secondary spectrum than that on the primary spectrum.  

The secondary-to-primary ratio can be calculated by simply taking the ratio of Eq. (9) to Eq. (7). For the case of no re-acceleration $\xi=0$, it can be checked that both Eqs. (7) and (9) reduce to the standard solution of pure-diffusion equation (see e.g., \citealp{Thoudam2008}), and also that the secondary-to-primary ratio becomes proportional to $1/D$ at high energies. Here again, a steeper re-acceleration index $s>q$ will produce an enhancement in the ratio at lower energies, and unlike in the case of primary spectra, a harder index $s=q$ will result into significant flattening of the ratio at high energies \citep{Wandel1987, Berezhko2003}. Thus, the effect of re-acceleration on cosmic-ray properties in the Galaxy depends strongly on the index of re-acceleration. In the present study, since we assume that re-acceleration is produced mainly by the interactions with old supernova remnants, we will only consider the case of $s>q$ with $s\gtrsim 4$. This value of $s$ corresponds to a Mach number of approximately $1.7$ for the shocks that re-accelerate cosmic rays.   

\section{Results and discussions}
For the present calculations, the inelastic interaction cross-section for cosmic-ray protons is taken from \citealp{Kelner2006}, and for heavier nuclei, the cross-sections are taken from \citealp{Letaw1983}. For cosmic-ray secondaries, we consider only the boron nuclei, and their production cross-section from the carbon and oxygen primaries are taken from \citealp{Heinbach1995}.  

For the interstellar matter density, we consider the averaged surface density on the Galactic disk within a radius equivalent to the halo height $H$. We take $H=5$ kpc for our study, and the averaged surface density of atomic hydrogen as $\bar{n}=7.24\times 10^{20}$ atoms cm$^{-2}$ \citep{Thoudam2013}. We further assume that the interstellar medium consists of $10\%$ helium.

\subsection{Effect of re-acceleration}  
Here, we demonstrate the effect of re-acceleration on the cosmic-ray energy spectra and the secondary-to-primary ratios. Figure 1 left panel shows an example of the proton energy spectrum (solid line), calculated using Eq. (7), decomposed into the re-accelerated (dashed line) and the normal (dotted-dashed line) components. For the demonstration, the calculation is performed at $z=0$, and assumes $D_0=5\times 10^{28}$ cm$^2$ s$^{-1}$, $\rho_0=3$ GV and $a=0.33$. The re-acceleration parameters are taken as $\eta=0.6$ and $s=4.0$, and  the source parameters as $q=2.3$ and $p_c=\infty$. For this particular set of model parameters, it is found that the re-accelerated component dominates up to $\sim 1$ TeV while above, the spectrum is dominated by the normal component. At  energies above $\sim 20$ GeV, the re-accelerated component is steeper following an index close to $\sim s+a=4.33$, compared to the normal component which has an index of $\sim q+a=2.63$. It can be noticed that this steep re-acceleration component produces a bump in the total spectrum below $\sim 1$ TeV resulting into spectral difference between GeV and TeV energy regions. The magnitude of the bump depends on the amount of re-acceleration which is related to the value of $\eta$. Choosing lower values of $\eta$ will decrease the re-accelerated component, while at the same time increases the contribution of the normal component, thus reducing the bump in the total spectrum. In addition, the effect of re-acceleration also depends on the choice of $q$. A larger $q$ generates a steeper cosmic-ray spectrum in the Galaxy, thereby providing a higher number of low-energy particles for re-acceleration, and this increases the re-acceleration component.   
\begin{figure}
\centering
\includegraphics*[width=0.34\textwidth,angle=-90,clip]{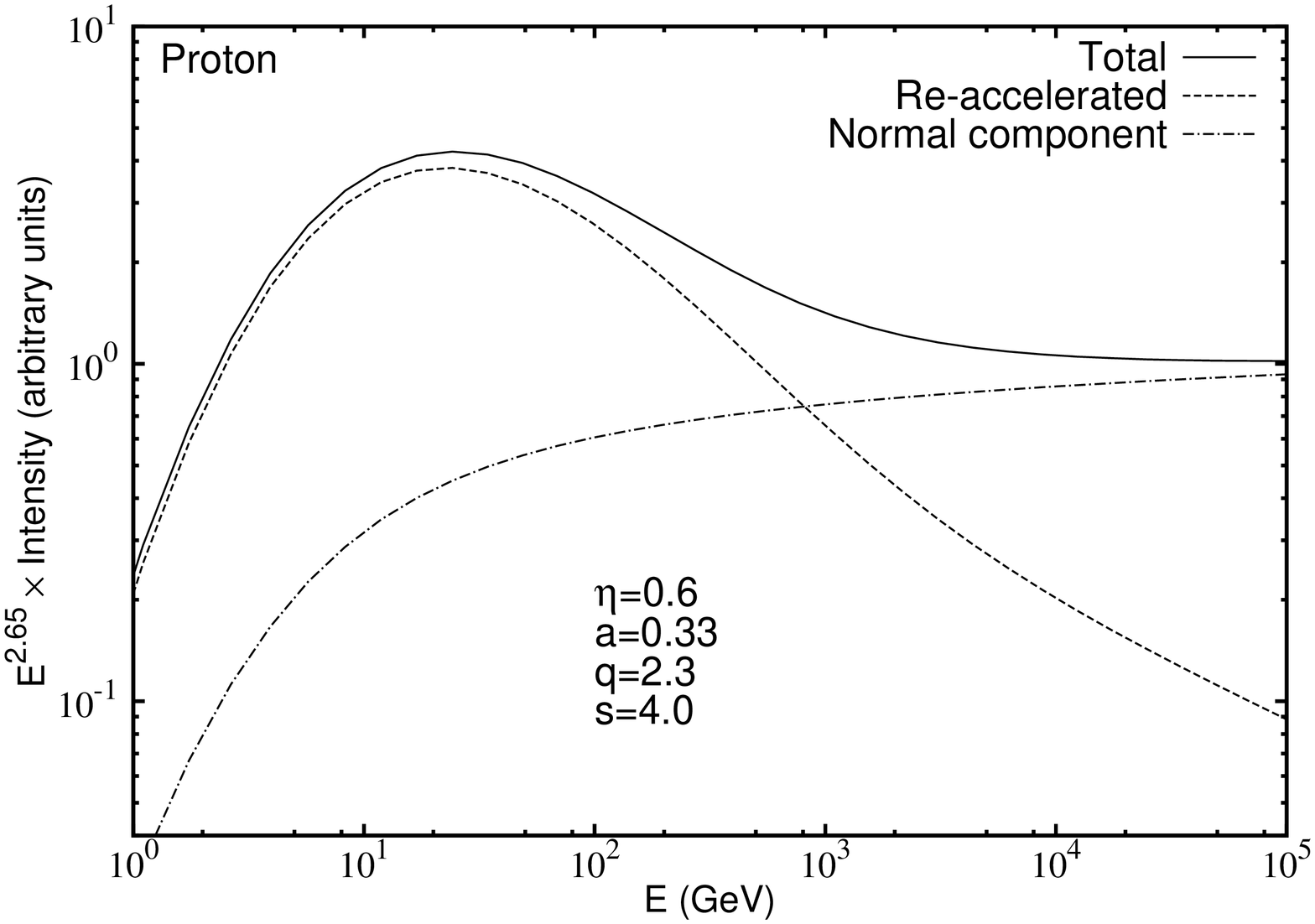}
\includegraphics*[width=0.34\textwidth,angle=-90,clip]{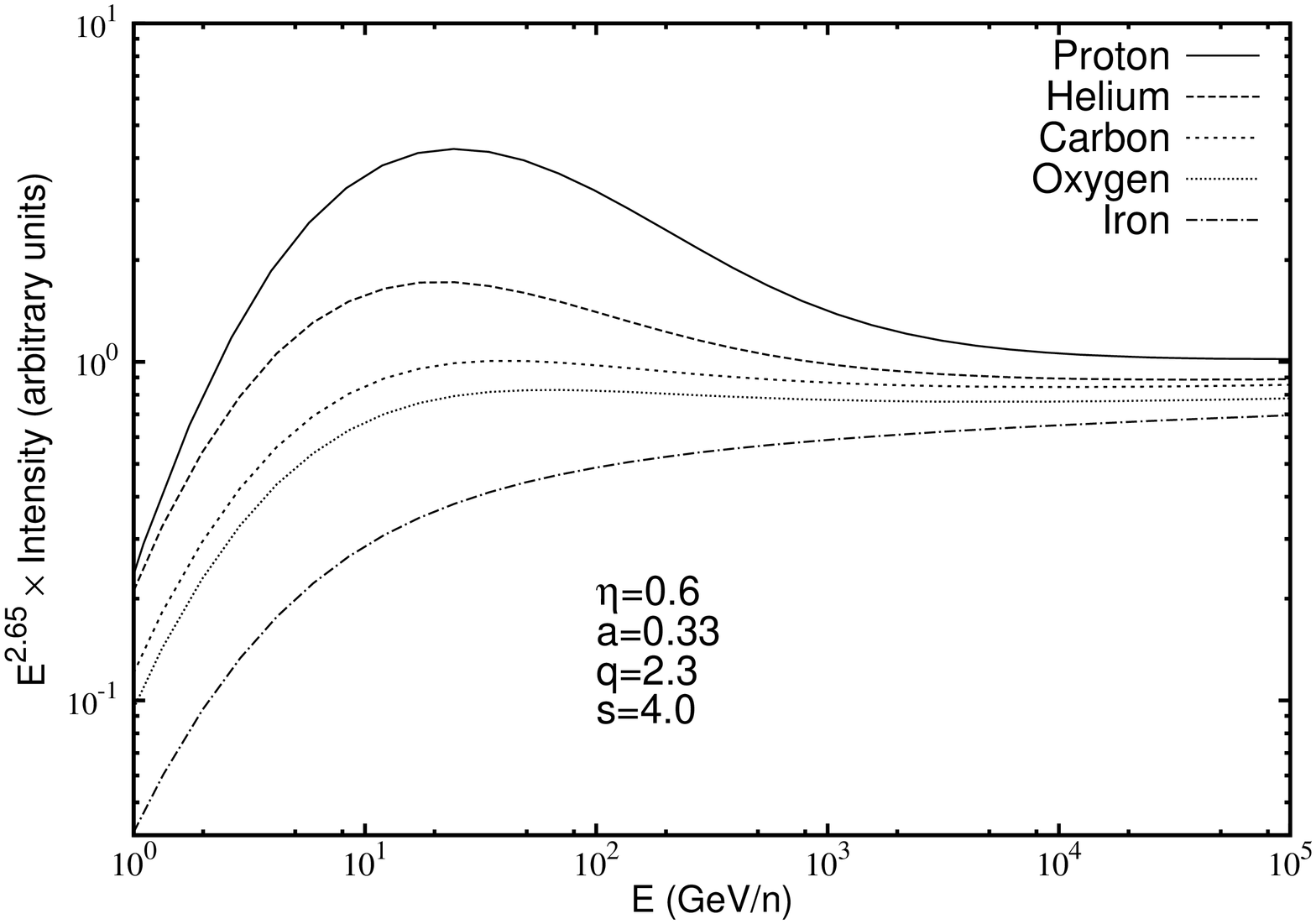}
\caption{\label {fig1} Left: Proton spectrum showing the re-accelerated and the normal components. Right: Re-acceleration effect on different elements. Lines top to bottom: proton, helium, carbon, oxygen, and iron. Model parameters used: $\eta=0.6$, $D_0=5\times 10^{28}$ cm$^2$ s$^{-1}$, $\rho_0=3$ GV, $a=0.33$, $q=2.3$, $s=4.0$, and $p_c=\infty$.}
\end{figure}

The effect of re-acceleration on different types of nuclei (proton, helium, carbon, oxygen, and iron) is shown in Figure 1 right panel. The results shown are for the same set of model parameters used in Figure 1 left panel. It can be seen that the re-acceleration effect is more prominent for lighter nuclei than for heavier nuclei. It is maximal for protons, and minimal for iron nuclei. In other words, in the present model, light nuclei will show larger GeV-TeV spectral differences than heavy nuclei. For heavier nuclei, due to their larger interaction cross-section, inelastic collisions dominate over re-acceleration. But, for lighter nuclei, such as proton and helium, for which the inelastic cross-sections are relatively small, they can be efficiently re-accelerated during their residence time in the Galaxy. This decreasing effect of re-acceleration for an increasing elemental mass is expected only in the present model. 

\begin{figure}
\centering
\includegraphics*[width=0.34\textwidth,angle=-90,clip]{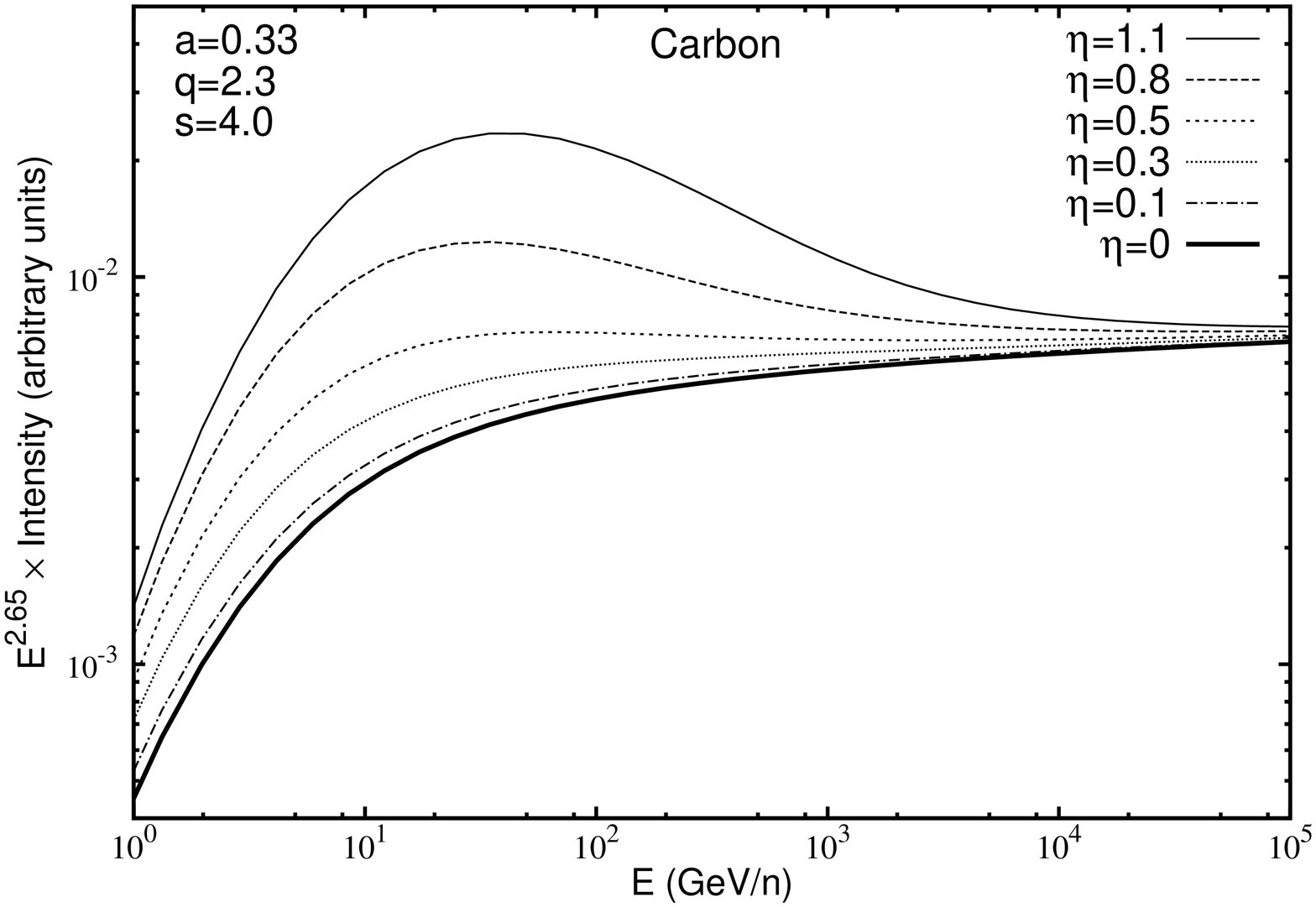}
\includegraphics*[width=0.34\textwidth,angle=-90,clip]{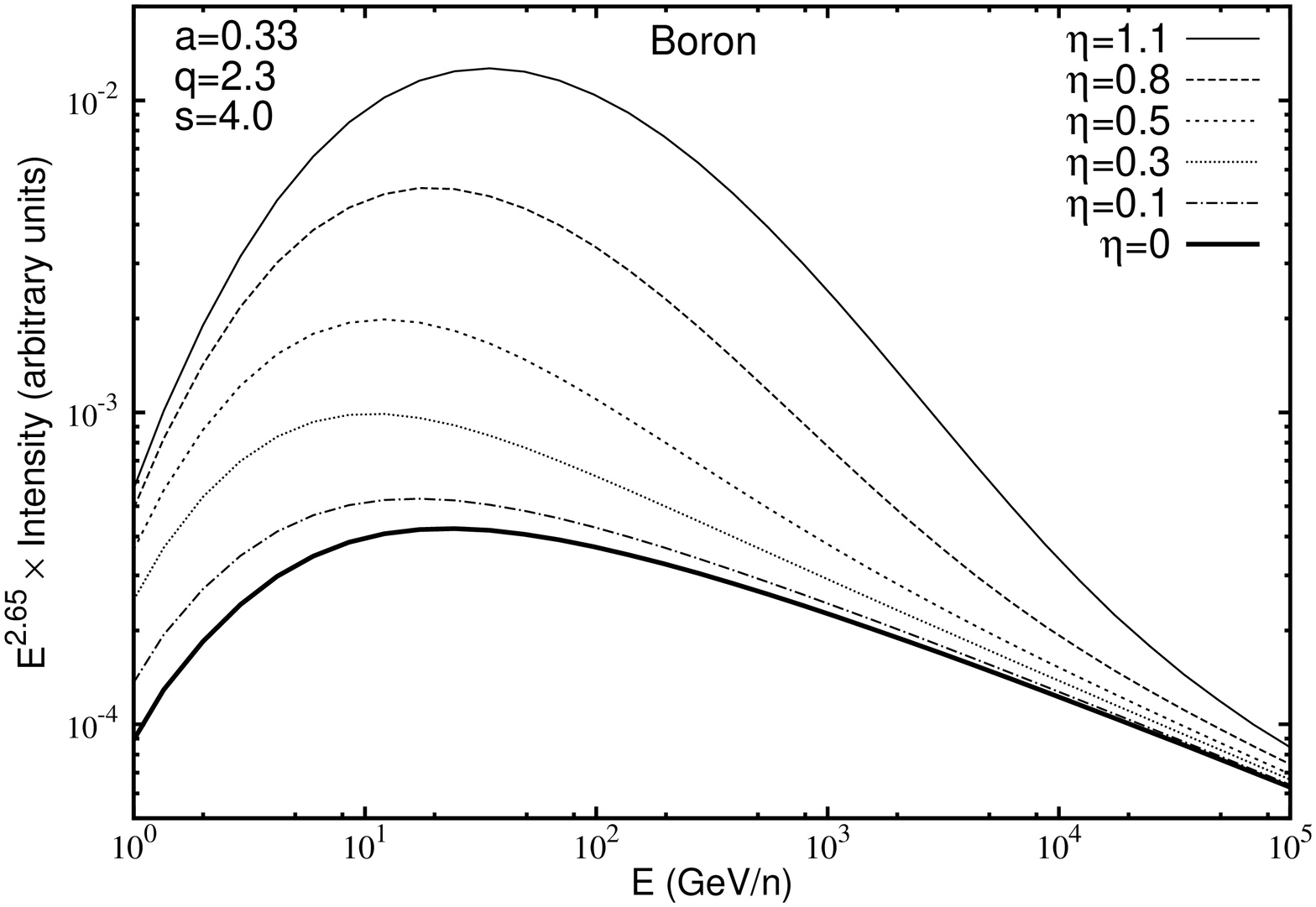}
\caption{\label {fig1} Carbon (left) and boron (right) spectra for $\eta=(0, 0.1, 0.3, 0.5, 0.8, 1.1)$. Other model parameters remain the same as in Figure 1.}
\end{figure}

In Figure 2, the carbon (left panel) and boron (right panel) energy spectra, calculated using Eqs. (7) and (9) respectively, are shown as function of kinetic energy/nucleon. The results are calculated at $z=0$, and correspond to different levels of re-acceleration given by the parameter $\eta$ taken in the range of $\eta=0-1.1$. The thick solid line represents no re-acceleration, i.e. $\eta=0$, and the thin lines correspond to some finite amount of re-acceleration with larger $\eta$ corresponding to higher level of re-acceleration. Other model parameters are taken to be the same as in Figure 1. For the calculation of the  boron spectrum, the source parameters are taken to be the same for both the carbon and oxygen nuclei. It can be noticed that as $\eta$ takes larger values, the spectral bump  due to re-acceleration increases as expected, and also at the same time the re-acceleration effect becomes extended to higher and higher energies. For $\eta=1.1$, the maximum value considered here, the effect is observed up to a few TeV/n in the carbon spectrum.

Compared to the carbon spectrum, the re-acceleration effect is found to be more prominent, and also more extended in energy, in the case of boron. There is some mild effect due to the slightly lower inelastic cross-section of boron nuclei with respect to the carbon that makes the re-acceleration more efficient for boron, but this effect is negligible. The main effect, as mentioned in Section 2, is due to the contribution from the re-acceleration of boron which are produced by the re-accelerated component of the primary carbon and oxygen nuclei. Moreover, the non re-accelerated (normal) component of boron given by the first term of Eq. (9) is steeper than that of the carbon due to the energy dependent escape of cosmic rays from the Galaxy. It may be noted that the spectrum of the normal secondary component is steeper than that of the primaries by the index of diffusion. This allows the re-accelerated component to dominate up to higher energies in the case of boron.

\begin{figure}
\centering
\includegraphics*[width=0.34\textwidth,angle=-90,clip]{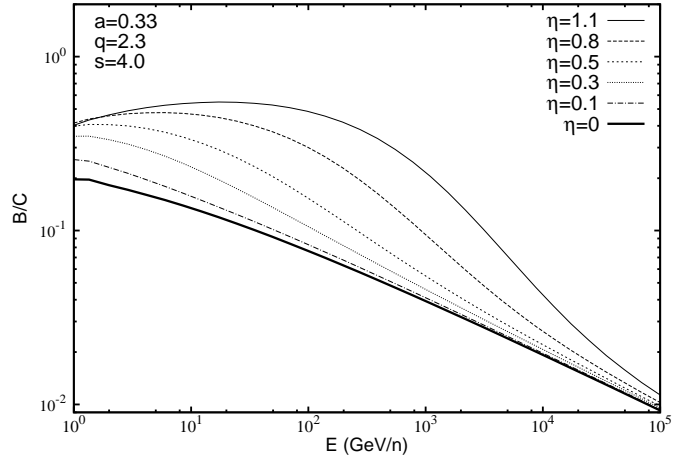}
\caption{\label {fig1} Boron-to-carbon ratio for $\eta=(0, 0.1, 0.3, 0.5, 0.8, 1.1)$. Other model parameters are the same as in Figure 1.}
\end{figure}
Figure 3 shows the boron-to-carbon ratio for the different values of $\eta$. The model parameters and the line representations remain the same as in Figure 2. Similar effects observed in the energy spectra shown in Figure 2 are also observed in the ratio. In the model without re-acceleration ($\eta=0$), the ratio follows an inverse relation with the diffusion coefficient, and hence, the slope of the ratio follows the inverse of the diffusion index as $E^{-a}$ (see thick solid line in Figure 3). When comparing the result for $\eta=0$ with the results obtained for $\eta>0$, it is clear that in the re-acceleration model, the secondary-to-primary ratio does not represent a direct measure of the cosmic-ray diffusion coefficient in the Galaxy as in pure diffusion models. The ratio also depends on the re-acceleration parameters such as the efficiency of re-acceleration and the spectral index of the re-accelerated particles $s$. Moreover, the ratio depends weakly on the primary source parameters such as $q$ and $f$, unlike in the pure diffusion models where the ratio is almost independent of the source parameters.

\subsection{Comparison with the data}
For the rest of the study, we take the size of the source distribution $R=20$ kpc, the proton high-momentum cut-off $p_c=1$ PeV/c, and the supernova explosion rate as $\bar{\nu}=25$ SNe Myr$^{-1}$ kpc$^{-1}$. The latter corresponds to a rate of $\sim 3$ SNe per century in the Galaxy. The cosmic-ray propagation parameters $(D_0, \rho_0, a)$, the re-acceleration parameters $(\eta, s)$ and the source parameters $(q, f)$ are taken as model parameters. They are derived from the measured cosmic-ray data.

\begin{figure}
\centering
\includegraphics*[width=0.34\textwidth,angle=-90,clip]{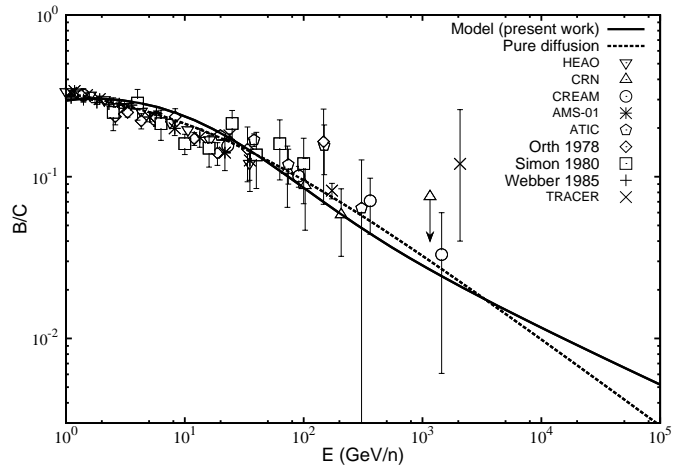}
\caption{\label {fig1} Boron-to-carbon ratio. {\it Solid line}: Our present result for maximum re-acceleration. {\it Dashed line}: Best-fit result for pure diffusion model \citep{Thoudam2013}. Model parameters used: $\eta=1.02$, $D_0=9\times 10^{28}$ cm$^2$ s$^{-1}$, $\rho_0=3$ GV, $a=0.33$, $q_C=2.24$, $q_O=2.26$, $s=4.5$, $p_c=1$ PeV/c, $f_C=0.024\%$, $f_O=0.025\%$, $\bar{\nu}=25$ SNe Myr$^{-1}$ kpc$^{-2}$ and $\phi=450$ MV. Data: HEAO \citep{Engelmann1990}, CRN \citep{Swordy1990}, CREAM \citep{Ahn2008}, AMS-01 \citep{Aguilar2010}, ATIC \citep{Panov2008}, \citealp{Orth1978}, \citealp{Simon1980}, \citealp{Webber1985}, and TRACER \citep{Obermeier2011}.}
\end{figure}

We first determine $(D_0, \rho_0, a, \eta, s)$ based on the measurements of the boron-to-carbon ratio, and the spectra for the  carbon, oxygen, and boron nuclei simultaneously. Their values are found to be $D_0=9\times 10^{28}$ cm$^2$ s$^{-1}$, $\rho=3$ GV, $a=0.33$, $\eta=1.02$, and $s=4.5$. These values correspond to the maximum amount of re-acceleration permitted by the available boron-to-carbon data, while at the same time produces a reasonable good fit to the measured carbon, oxygen, and boron energy spectra simultaneously. Figure 4 shows the result on the boron-to-carbon ratio (solid line) along with the measurement data. The data are from HEAO \citep{Engelmann1990}, CRN \citep{Swordy1990}, CREAM \citep{Ahn2008}, AMS-01 \citep{Aguilar2010}, ATIC \citep{Panov2008}, \citealp{Orth1978}, \citealp{Simon1980}, \citealp{Webber1985}, and TRACER \citep{Obermeier2011}. For comparison, we have also shown the best-fit result for the case of pure diffusion (dashed line), i.e. no re-acceleration $(\eta=0)$, taken from \citealp{Thoudam2013}. The diffusion index of $a=0.33$ obtained in the present model is the same as that found in models of cosmic-ray distributed re-acceleration due to interstellar turbulence \citep{Strong1998}. However, the value of $D_0$ obtained here is slightly larger than that obtained in \citealp{Strong1998} which gave a value of $7.7\times 10^{28}$ cm$^2$ s$^{-1}$ for the same value of $H=5$ kpc.

\begin{figure}
\centering
\includegraphics*[width=0.34\textwidth,angle=-90,clip]{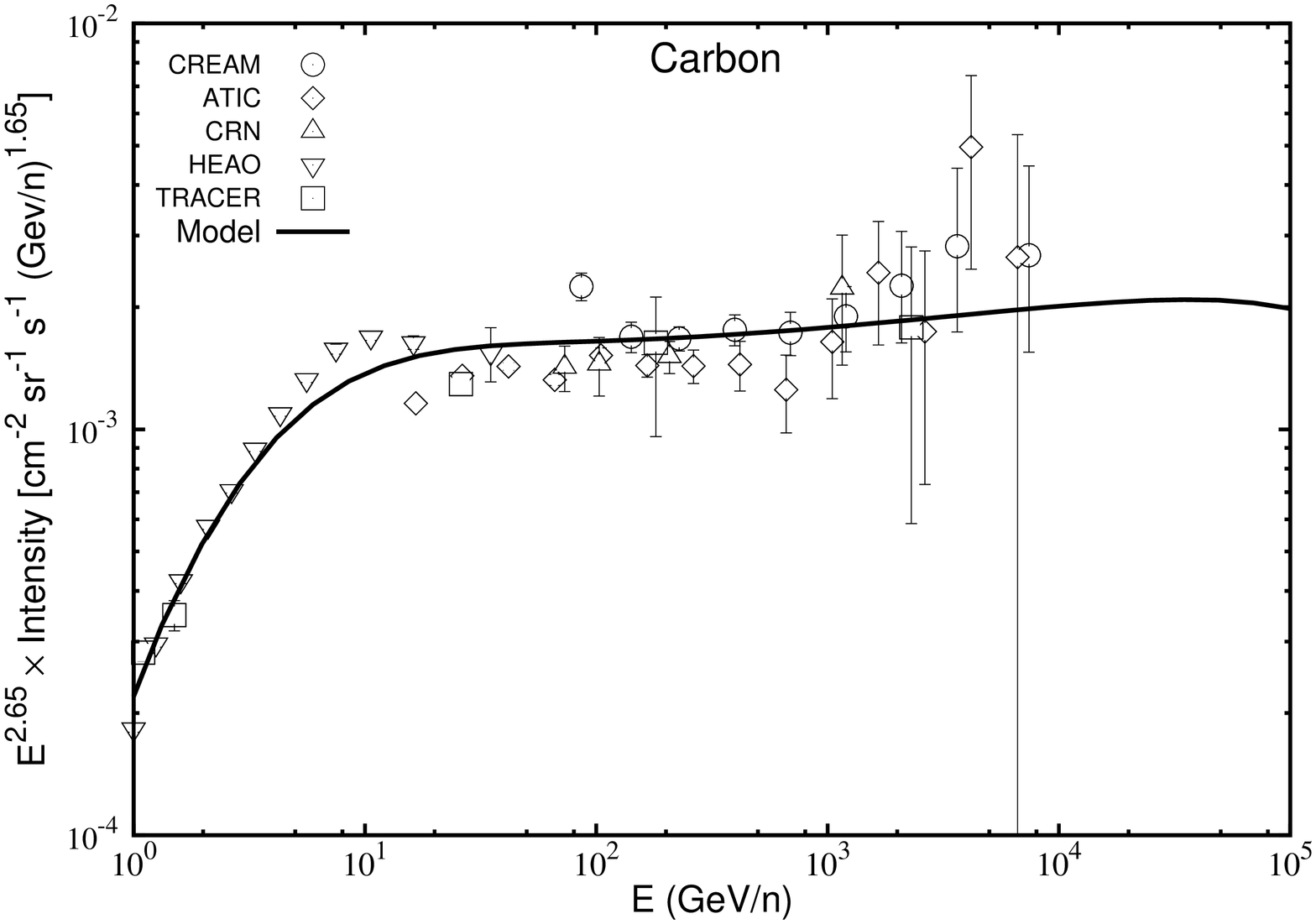}\includegraphics*[width=0.34\textwidth,angle=-90,clip]{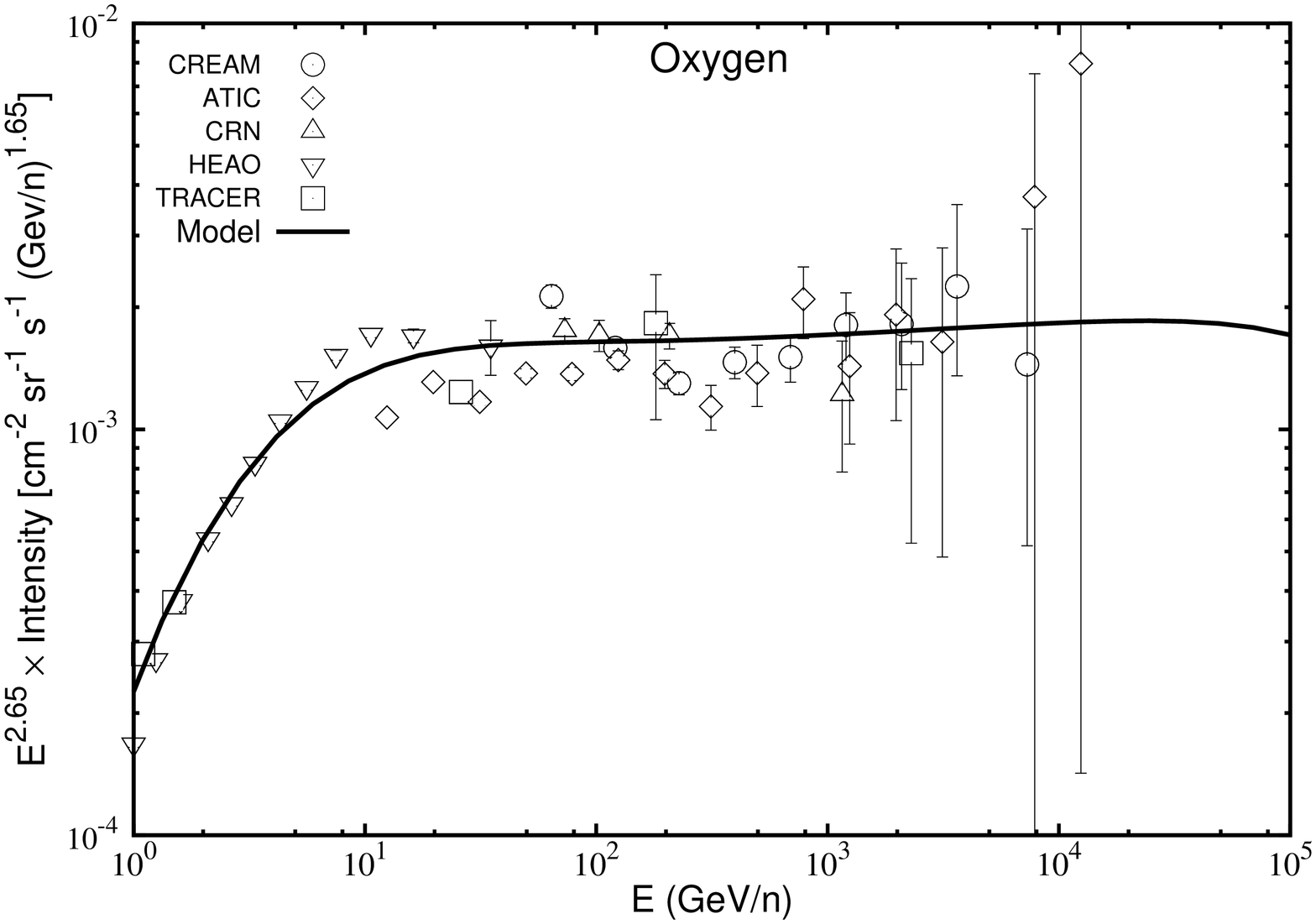}\\
\includegraphics*[width=0.34\textwidth,angle=-90,clip]{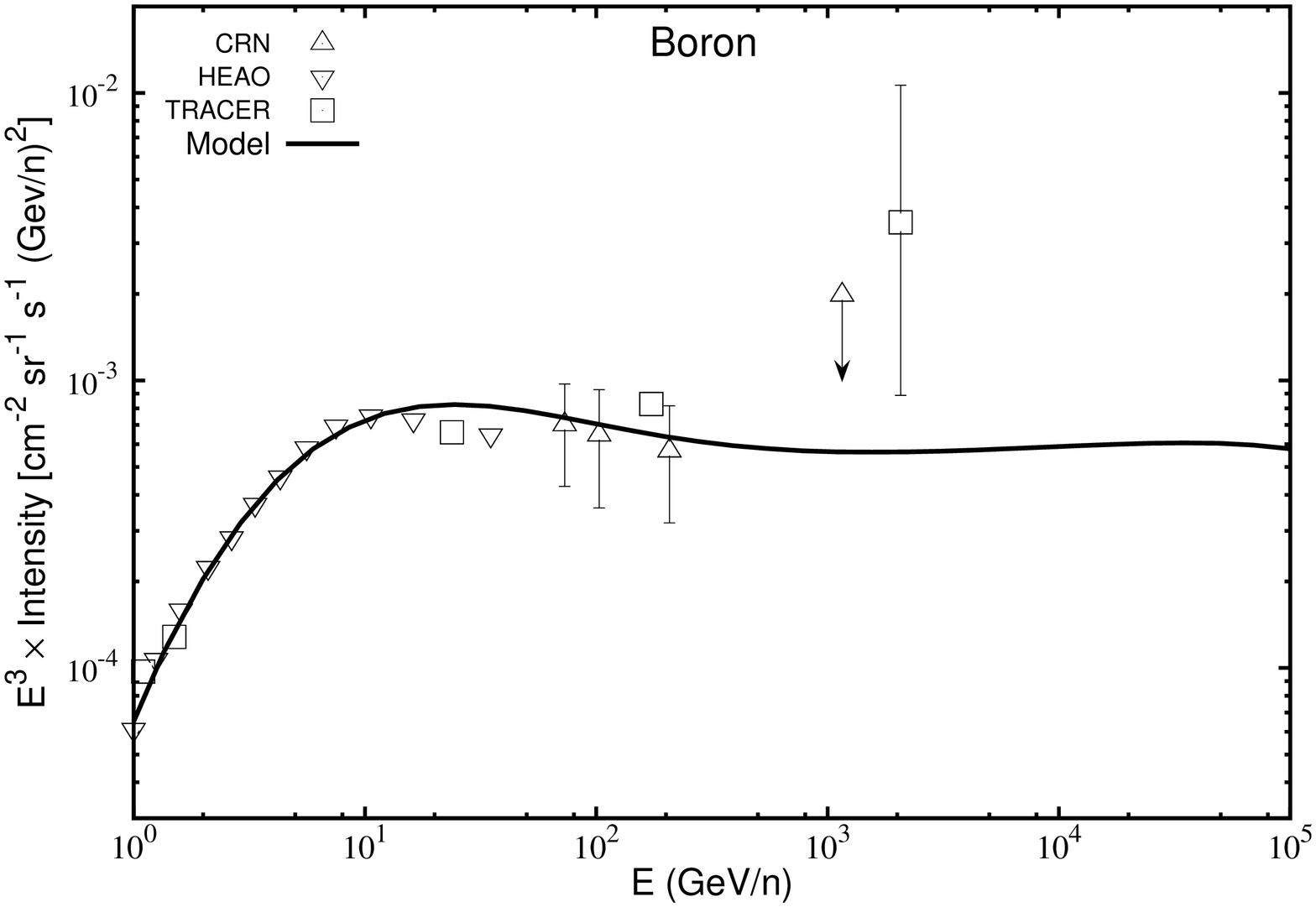}
\caption{\label {fig1} Carbon, oxygen and boron energy spectra. {\it Solid line}: Our calculation. Model parameters are the same as in Figure 4. Data: CREAM \citep{Ahn2009}, ATIC \citep{Panov2007}, CRN \citep{Mueller1991, Swordy1990}, HEAO \citep{Engelmann1990}, and TRACER \citep{Obermeier2011}.}
\end{figure}

The corresponding results on carbon, oxygen, and boron energy spectra are shown in Figure 5. The solid line represents our results, and the data are taken from CREAM \citep{Ahn2009}, ATIC \citep{Panov2007}, CRN \citep{Mueller1991, Swordy1990}, HEAO \citep{Engelmann1990}, and TRACER \citep{Obermeier2011}. The carbon and oxygen source parameters used are $(q_C=2.24, f_C=0.024\%)$, and $(q_O=2.26$, $f_O=0.025\%)$ respectively, where the $f$'s are given in units of $10^{51}$ ergs. Our calculation assumes a force-field solar modulation parameter of $\phi=450$ MV. Our model does not produce  a  significant spectral hardening for both the carbon and the oxygen spectra although a slight hardening is noticed above $\sim 100$ GeV/n in the case of carbon. Given the large uncertainties on the measurements above $\sim 1$ TeV/n, our predictions are found to be in agreement with the data up to $\sim 10$ TeV/n. For boron nuclei, our model predicts a noticeable spectral hardening above $\sim 500$ GeV/n, which is due to the increase in the effect of re-acceleration on the secondaries relative to the primaries as discussed above. A Similar effect is also visible in the boron-to-carbon ratio shown in Figure 4 (see solid line). Our prediction is different from that of both the pure diffusion, and the distributed re-acceleration models, which predict spectra close to a pure power-law at energies above $\sim 20-30$ GeV/n. Although our model does not effectively reproduce the highest data point measured by the TRACER experiment, it seems to be consistent with the apparent spectral hardening indicated by the measurement. It can be mentioned that such a hardening in the secondary spectrum can also be attributed to additional components of cosmic-ray secondaries which might exist in the Galaxy and have not been considered in the present model, such as those produced by the interaction of primaries inside the sources or those which results from the re-acceleration of background secondaries by strong shocks (see e.g. \citealp{Berezhko2003, Wandel1987}). Such secondaries, although expected to represent a small fraction at low energies, might become important at high energies above $\sim 1$ TeV/n because of their harder energy spectrum compared to the secondaries produced in the interstellar medium.  

Using the same values of $(D_0, \rho_0, a, \eta, s)$ obtained above, we calculate the spectra for the proton and helium nuclei. The results are shown in Figure 6, where the left panel represents proton and the right panel represents helium. The lines represent our results, and the data are taken from the CREAM \citep{Yoon2011}, ATIC \citep{Panov2007}, AMS-01 \citep{Alcaraz2000, Aguilar2002}, and PAMELA \citep{Adriani2011} experiments. The source parameters used are $(q_p=2.21, f_p=6.95\%)$ for protons, and $(q_{He}=2.18, f_{He}=0.79\%)$ for helium, and we use the same solar modulation parameter as given above. It can be seen that our results are in good agreement with the measured data, and explain the observed spectral anomaly between the GeV and TeV energy regions. Below $\sim 200$ GeV/n, our model spectrum is dominated by the re-accelerated component while above, it is dominated by the normal component. The spectral roll-off above $\sim 10^5$ GeV/n is due to the assumed exponential cut-off of the source spectrum at $p_c$ which we keep fixed at $10^6$ GeV/$c$ for protons.  

\begin{figure}
\centering
\includegraphics*[width=0.34\textwidth,angle=-90,clip]{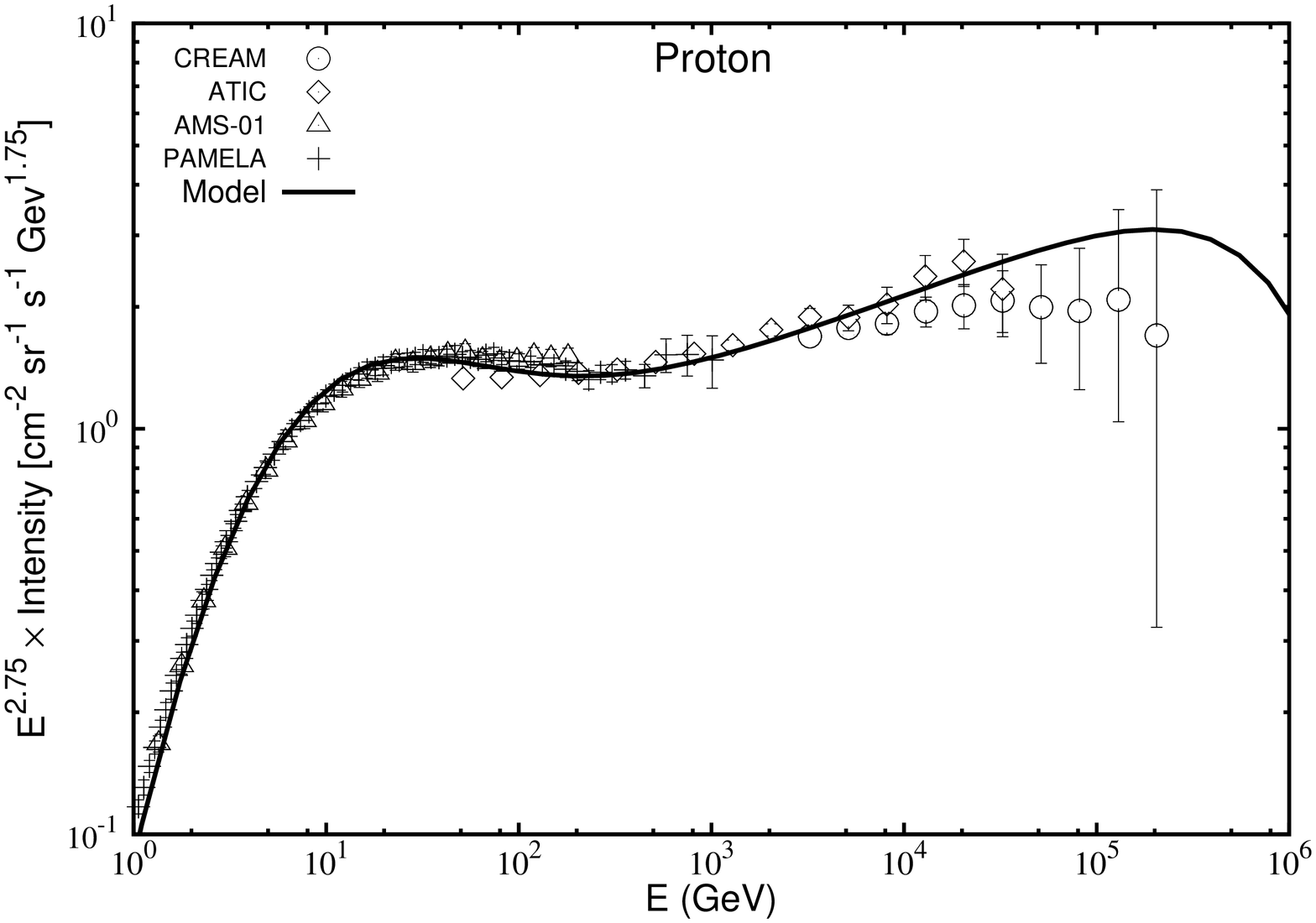}
\includegraphics*[width=0.34\textwidth,angle=-90,clip]{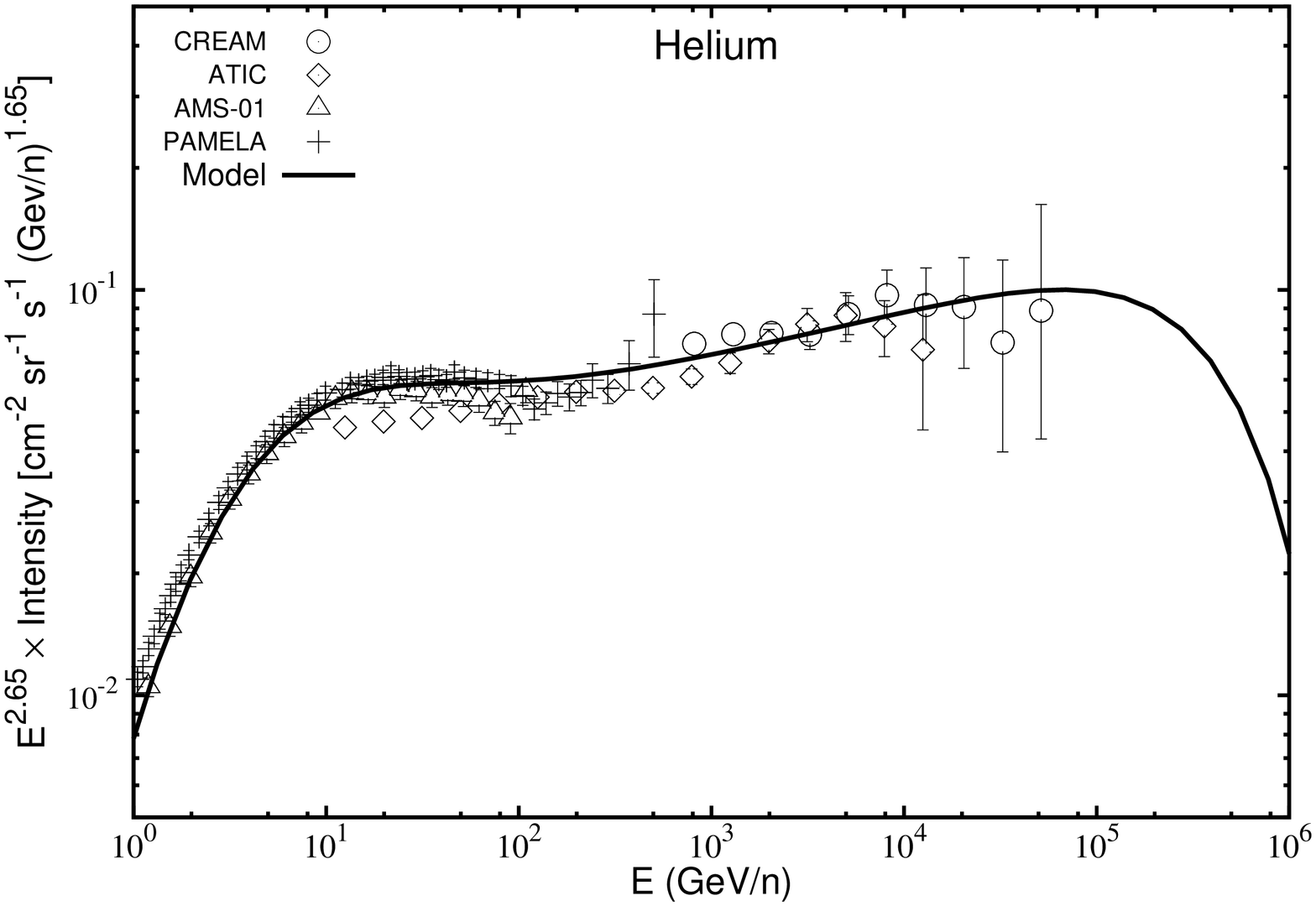}
\caption{\label {fig1} Result for protons (left) and helium nuclei (right). {\it Solid line}: Our calculation. Model parameters used: $q_P=2.21$, $q_{He}=2.18$, $f_P=6.95\%$, $f_{He}=0.79\%$. The propagation and the re-acceleration model parameters $(D_0, \rho_0, a, \eta, s)$ are the same as in Figure 4. Data: CREAM \citep{Yoon2011}, ATIC \citep{Panov2007}, AMS-01 \citep{Alcaraz2000, Aguilar2002}, and PAMELA \citep{Adriani2011}.}
\end{figure}

Our result shows that the re-acceleration effect is stronger in the case of protons resulting into more prominent spectral differences in the GeV-TeV region for protons than for helium. This is partly due to the effect of larger inelastic collision losses for helium nuclei than protons as shown in Figure 1 (right panel). For the present set of model parameters, there is also an additional effect due to the steeper proton source index of $q_p=2.21$ compared to that of helium nuclei of $q_{He}=2.18$. Choosing a larger index produces a steeper spectrum of background cosmic rays in the Galaxy. This leads to two effects on the re-accelerated component: First, a larger number of low-energy background particles become available for re-acceleration, leading to an increase in the number of re-accelerated particles. And secondly, because now the normal component also becomes steeper, the contribution of the re-acelerated component becomes more extended to higher energies. Therefore, the re-acceleration effect turns out to be more prominent, and also somewhat more extended in energy for protons than for helium.
\begin{figure}
\centering
\includegraphics*[width=0.34\textwidth,angle=-90,clip]{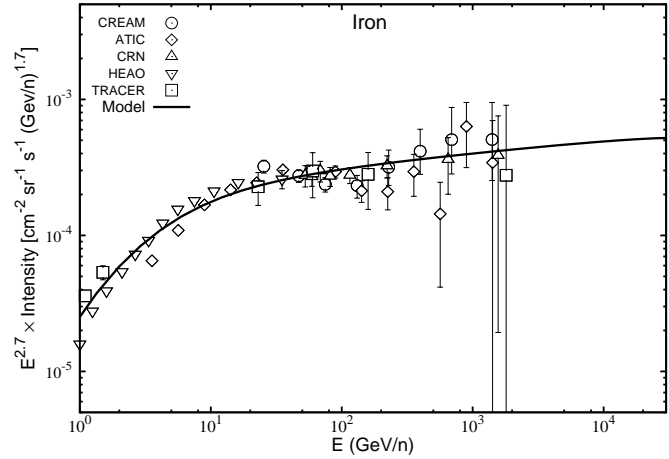}
\caption{\label {fig1} Result for iron nuclei. {\it Solid line}: Our calculation. Model parameters used: $q_{Fe}=2.28$,  $f_{Fe}=0.0049\%$. All other model parameters remain the same as in Figure 4. Data: CREAM \citep{Ahn2009}, ATIC \citep{Panov2007}, CRN \citep{Swordy1990}, HEAO \citep{Engelmann1990}, and TRACER \citep{Obermeier2011}.}
\end{figure}

For heavier nuclei for which the inelastic cross-sections are much larger, the re-acceleration effect is significantly less even for a steeper source spectrum. This is demonstrated in Figure 7 with our result on the iron nuclei. The calculation assumes the source parameters to be $q_{Fe}=2.28$ and $f_{Fe}=4.9\times 10^{-3}\%$ to reproduce the measured spectrum. The propagation and the re-acceleration model parameters $(D_0, \rho_0, a, \eta, s)$ are taken to be the same as given in Figure 4. It can be seen that although the source spectrum is taken to be much steeper, the re-acceleration effect is hard to notice in Figure 7, and the model spectrum above $\sim 20$ GeV/n follows approximately a single power-law, unlike the proton and helium spectra. Thus, our present model predicts a mass dependent spectral hardening, which can be used to differentiate it from other models in future. Furthermore, it can be mentioned that in our model, such a spectral hardening is not expected for electrons as they suffer severe radiative losses which will dominate the re-acceleration effect even at few GeV energies.

\section{Conclusion}
In short, we conclude that the spectral anomaly of cosmic rays at GeV-TeV energies, observed for the proton and helium nuclei by recent experiments, can be an effect of re-acceleration of cosmic rays by weak shocks associated with old supernova remnants in the Galaxy. The re-acceleration effect is shown to be important  for light nuclei such as proton and helium, and negligible for heavier nuclei such as iron. Our prediction of decreasing effect of re-acceleration with the increase in the elemental mass can be checked by future sensitive measurements of heavier nuclei at TeV/n energies. The re-acceleration  effect is expected to be negligible for electrons.

\section*{Acknowledgements}
The authors wish to thank Prof. Reinhard Schlickeiser for his critical comments and suggestions on the mathematical derivation given in the Appendix.


\appendix
\section{Derivation of Green's function for Eq. (1)}
The Green's function $G(\textbf{r},\textbf{r}^\prime,p,p^\prime)$ for Eq. (1) satisfies
\begin{equation}
\nabla\cdot(D\nabla G)-\left[\bar{n} v\sigma+\xi\right]\delta(z)G+\left[\xi sp^{-s}\int^p_{p_0}du\;G(u)u^{s-1}\right]\delta(z)=-\delta(\textbf{r}-\textbf{r}^\prime)\delta(p-p^\prime)
\end{equation}
where $G(u)\equiv G(\textbf{r},\textbf{r}^\prime,u,p^\prime)$. In rectangular coordinates, and assuming the sources to be on the Galactic plane, i.e., $z^\prime=0$, we can write the above equation as
\begin{equation}
D\frac{\partial^2 G}{\partial x^2}+D\frac{\partial^2 G}{\partial y^2}+D\frac{\partial^2 G}{\partial z^2}-\left[\bar{n} v\sigma+\xi\right]\delta(z)G+\left[\xi sp^{-s}\int^p_{p_0}du\;G(u)u^{s-1}\right]\delta(z)=-\delta(x-x^\prime)\delta(y-y^\prime)\delta(z)\delta(p-p^\prime)
\end{equation}
After taking Fourier transform of Eq. (A.2) with respect to $x$ and $y$, we have
\begin{equation}
-Dk^2\bar{G}+D\frac{\partial^2 \bar{G}}{\partial z^2}-\left[\bar{n} v\sigma+\xi\right]\delta(z)\bar{G}+\left[\xi sp^{-s}\int^p_{p_0}du\;\bar{G}(u)u^{s-1}\right]\delta(z)=-\exp(ik_x x^\prime+ik_y y^\prime)\delta(z)\delta(p-p^\prime)
\end{equation}
where $k^2=k_x^2+k_y^2$, and
\begin{equation}
\bar{G}(k_x,x^\prime,k_y,y^\prime,z,p,p^\prime)=\int_{-\infty}^\infty dx\int_{-\infty}^\infty dy\;G(x,x^\prime,y,y^\prime,z,p,p^\prime)\exp(ik_xx+ik_yy)\nonumber
\end{equation}
For $z\neq 0$, Eq. (A.3) reduces to the following simple differential equation:
\begin{equation}
-Dk^2\bar{G}+D\frac{\partial^2 \bar{G}}{\partial z^2}=0\nonumber
\end{equation}
Solving the above equation by using the boundary condition that the particle density, and hence $\bar{G}$, vanishes at $z=\pm H$, the solution of Eq. (A.3) for regions above $(z>0)$ and below $(z<0)$ the Galactic plane is obtained as,
\begin{equation}
\bar{G}(k_x,x^\prime,k_y,y^\prime,z,p,p^\prime)=\bar{G}(0)\frac{\sinh\left[k(H-|z|)\right]}{\sinh(kH)}
\end{equation}
where $\bar{G}(0)\equiv \bar{G}(k_x,x^\prime,k_y,y^\prime,z=0,p,p^\prime)$. The function $\bar{G}(0)$ can be determined using the continuity relation at $z=0$ as follows. Integrating Eq. (A.3) over $z$ around $z=0$, we get
\begin{equation}
\left. D\frac{\partial\bar{G}}{\partial z}\right\vert_{z=0+}-\left. D\frac{\partial\bar{G}}{\partial z}\right\vert_{z=0-}-\left[\bar{n} v\sigma+\xi\right]\bar{G}(0)+\left[\xi sp^{-s}\int^p_{p_0}du\;\bar{G}(0,u)u^{s-1}\right]=-\exp(ik_x x^\prime+ik_y y^\prime)\delta(p-p^\prime)
\end{equation}
where $\bar{G}(0,u)\equiv \bar{G}(k_x,x^\prime,k_y,y^\prime,0,u,p^\prime)$. Substituting for $\partial \bar{G}/\partial z$ at $z=0\pm$, and rearranging the terms, Eq. (A.5) reduces into a first order linear differential equation in $p$ as
\begin{equation}
\frac{d\bar{G}(0)}{dp}+A(p)\bar{G}(0)=-B(p)
\end{equation}
where,
\begin{align}
& A(p)=\frac{s}{p}-\frac{\xi s}{pL(p)}+\frac{1}{L(p)}\frac{d}{dp}L(p)\nonumber\\
& L(p)=2D(p)k\coth(kH)+\bar{n}v(p)\sigma(p)+\xi\nonumber\\
& B(p)=-\frac{C(p)}{p^s}\frac{d}{dp}\left[ p^s \delta(p-p^\prime)\right]\nonumber\\
& C(p)=\frac{\exp(ik_x x^\prime+ik_y y^\prime)}{L(p)}
\end{align}
The integrating factor of Eq. (A.6) is:
\begin{equation}
I.F=\exp\left(\int^pA(u)du\right)\nonumber
\end{equation}
With this, the general solution of Eq. (A.6) is obtained as,
\begin{align}
\bar{G}(0)\times \exp\left(\int^pA(u)du\right)&= -\int^pdp_1 B(p_1) \exp\left(\int^{p_1}A(u)du\right) +I_0,\; \mathrm{where\;\textit{I}_0\;is\;the\; integration\;constant}\nonumber\\
&=-\int^pdp_1 \frac{-C(p_1)}{p_1^s} \frac{d}{dp_1}[p_1^s\delta(p_1-p^\prime)]\exp\left(\int^{p_1}A(u)du\right) +I_0\nonumber\\
&=-\int^pdp_1 E(p_1) \frac{d}{dp_1}[p_1^s\delta(p_1-p^\prime)]+I_0\tag{A.8}
\end{align}
where we have written,
\begin{equation}
E(p_1)=-\frac{C(p_1)}{{p_1}^s}\exp\left(\int^{p_1} A(u)du\right)\tag{A.9}
\end{equation}
The first term on the right hand side of Eq. (A.8) can be integrated by parts taking $E(p_1)$ as the first function and the derivative part as the second function as follows:
\begin{align}
\bar{G}(0)\times \exp\left(\int^pA(u)du\right)&=-E(p)p^s\delta(p-p^\prime)+\int^pdp_1 p_1^s\delta(p_1-p^\prime) \frac{d}{dp_1}E(p_1) +I_0\nonumber\\
&=-E(p)p^s\delta(p-p^\prime)+H[p-p^\prime]{p^\prime}^s \frac{d}{dp^\prime} E(p^\prime) +I_0\tag{A.10}
\end{align}
where $H[m]$ is the Heaviside step function which has the property $H(m)=1(0)$ for $m>0 (<0)$. Substituting for $E(p)$, which is defined by Eq. (A.9), into Eq. (A.10), we get,
\begin{equation}
\bar{G}(0)\times \exp\left(\int^pA(u)du\right)=C(p)\delta(p-p^\prime)\exp\left(\int^pA(u)du\right)+H[p-p^\prime]{p^\prime}^s \frac{d}{dp^\prime}E(p^\prime) + I_0\tag{A.11}
\end{equation} 
Imposing the boundary condition that there are no particles at $p=0$, Eq. (A.11) becomes,
\begin{equation}
0=C(0)\delta(0-p^\prime)\exp\left(\int^{p=0}A(u)du\right)+H[0-p^\prime]{p^\prime}^s \frac{d}{dp^\prime}E(p^\prime) + I_0\nonumber
\end{equation}
In the present study, as we assume that particles are injected with a finite momentum $p^\prime\geq p_0$ where $p_0$ is the low-momentum  cut-off we have introduced so as to approximate the ionization losses (see Section 2), the delta function and the Heaviside step function in the first two terms on the right hand side of the above equation becomes zero. This gives $I_0=0$, and the general solution of Eq. (A.6) becomes,
\begin{equation}
\bar{G}(0)=C(p)\delta(p-p^\prime)+H[p-p^\prime]{p^\prime}^s \frac{d}{dp^\prime}E(p^\prime)\exp\left(-\int^p A(u)du\right)\tag{A.12} 
\end{equation}
Proceeding further, we have,
\begin{align}
\int A(u)du&=\int du\left(\frac{s}{u}-\frac{s\xi }{uL(u)}+\frac{1}{L(u)}\frac{d}{du}L(u)\right)\nonumber\\
&=\ln u^s+\xi s\int I(u)du +\ln L(u),\; \mathrm{where\;we\;have\;written}\;I(u)=-\frac{1}{uL(u)}\nonumber
\end{align}
Therefore, we can write
\begin{align}
\exp\left(\int A(u)du\right)&=u^s L(u)\exp\left(\xi s\int I(u)du\right)\tag{A.13}
\end{align}
Then, $E(p^\prime)$ defined by Eq. (A.9) becomes,
\begin{align}
E(p^\prime)=-\frac{C(p^\prime)}{{p^\prime}^s}{p^\prime}^s L(p^\prime)\exp\left(\xi s\int^{p^\prime} I(u)du\right)\tag{A.14}
\end{align}
Substituting for $C(p^\prime)$ as defined by Eq. (A.7) into Eq. (A.14), we get
\begin{align}
E(p^\prime)=-\exp(ik_x x^\prime+ik_y y^\prime)\exp\left(\xi s\int^{p^\prime} I(u)du\right)\tag{A.15}
\end{align}
Differentiating Eq. (A.15) with respect to $p^\prime$, we have
\begin{align}
\frac{d}{dp^\prime}E(p^\prime)&=-\exp(ik_x x^\prime+ik_y y^\prime)\exp\left(\xi s\int^{p^\prime} I(u)du\right)\frac{d}{dp^\prime}\left(\xi s\int^{p^\prime} I(u)du\right)\nonumber\\
&=-\exp(ik_x x^\prime+ik_y y^\prime)\exp\left(\xi s\int^{p^\prime} I(u)du\right)\xi s I(p^\prime)\nonumber\\
&=\exp(ik_x x^\prime+ik_y y^\prime)\exp\left(\xi s\int^{p^\prime} I(u)du\right)\frac{\xi s}{p^\prime L(p^\prime)}\tag{A.16}
\end{align}
where we have substituted $I(p^\prime)=-\frac{1}{p^\prime L(p^\prime)}$ in the last expression. Analogous to Eq. (A.13), we also obtain,
\begin{align}
\exp\left(-\int^p A(u)du\right)&=\frac{1}{p^s L(p)}\exp\left(-\xi s\int^p I(u)du\right)\tag{A.17}
\end{align}
Substituting Eqs. (A.16) and (A.17) into Eq. (A.12), we get
\begin{align}
\bar{G}(0)&=C(p)\delta(p-p^\prime)+H[p-p^\prime]{p^\prime}^s \exp(ik_x x^\prime+ik_y y^\prime)\frac{\xi s}{p^\prime L(p^\prime)} \exp\left(\xi s\int^{p^\prime} I(u)du\right) \frac{1}{p^s L(p)}\exp\left(-\xi s\int^p I(u)du\right)\nonumber\\
&=C(p)\delta(p-p^\prime)+H[p-p^\prime]\frac{\exp(ik_x x^\prime+ik_y y^\prime)}{L(p)}\frac{\xi s{p^\prime}^{s-1}}{p^s L(p^\prime)} \exp\left(\xi s\int^{p^\prime}_p I(u)du\right)\nonumber\\
&=\frac{\exp(ik_x x^\prime+ik_y y^\prime)}{L(p)}\left[\delta(p-p^\prime)+H[p-p^\prime]\frac{\xi s{p^\prime}^{s-1}}{p^s L(p^\prime)} \exp\left(\xi s\int^{p^\prime}_p I(u)du\right)\right],\; \mathrm{where\;we\;have\;substituted\;\textit{C(p)}\; from\;Eq.\;(A.7)}\nonumber\\
&=\exp(ik_x x^\prime+ik_y y^\prime) F(p,p^\prime)\tag{A.18}
\end{align}
where we have written,
\begin{align}
F(p,p^\prime)=\frac{1}{L(p)}\left[\delta(p-p^\prime)+H[p-p^\prime]\frac{\xi s {p^\prime}^{s-1}}{p^s L({p^\prime})} \exp\left(\xi s\int^{p^\prime}_p I(u) du\right)\right]\nonumber
\end{align}
Substituting for $\bar{G}(0)$ from Eq. (A.18) into Eq. (A.4), the Green's function for Eq. (1) can be obtained using the relation,
\begin{equation}
G(x,x^\prime,y,y^\prime,z,p,p^\prime)=\frac{1}{(2\pi)^2}\int_{-\infty}^\infty dk_x\int_{-\infty}^\infty dk_y\;\bar{G}(k_x,x^\prime,k_y,y^\prime,z,p,p^\prime)\exp(-ik_xx-ik_yy)\tag{A.19}
\end{equation}
In cylindrical coordinates $(r,r^\prime,z,z^\prime)$, where $x-x^\prime=(r-r^\prime)\cos\theta$, $y-y^\prime=(r-r^\prime)\sin\theta$, and $k_x=k\cos\phi$, $k_y=k\sin\phi$, we obtain
\begin{equation}
G(r,r^\prime,z,p,p^\prime)=\frac{1}{2\pi}\int^\infty_0 kdk\; F(p,p^\prime)\times\frac{\sinh\left[k(H-z)\right]}{\sinh(kH)}\times \mathrm{J_0}\left[k(r-r^\prime)\right]\tag{A.20}
\end{equation}
where $\mathrm{J_0}$ is a Bessel function of order 0. 

\end{document}